\documentclass[journal]{IEEEtran}

\usepackage{times}

\usepackage{algorithm}
\usepackage{algorithmic}
\usepackage{multirow}
\usepackage{amsmath}
\usepackage{bm}
\usepackage{xcolor}
\usepackage{subfigure}
\usepackage{url}
\usepackage{graphicx}
\graphicspath{ {./figs/} }

\newcounter{numberlistc}
\newenvironment{numberlist}
    {   \setcounter{numberlistc}{0}
        \begin{list}{\arabic{numberlistc}.}
        {\usecounter{numberlistc}
        \setlength{\parsep}{0pt}
        \setlength{\topsep}{3pt}
        \setlength{\itemsep}{0pt}}
        }{ \end{list} }
\newcounter{itemlistc}

\begin{document}
%\input replyJan20.tex
%\twocolumn
%date not printed
%\date{}
\title{GLU3.0:  Fast GPU-based Parallel Sparse LU Factorization for Circuit Simulation}

\author{Shaoyi~Peng,~\IEEEmembership{Student Member,~IEEE,}
  Sheldon~X.-D.~Tan,~\IEEEmembership{Senior Member,~IEEE, \\
  (Accepted for publication in IEEE Design \& Test in Feb 2020)}
  \thanks{This work is supported in part by NSF grant under
          No. CCF-1527324, and in part by NSF grant under
          No. CCF-1816361.}
\thanks{S. Peng and S. X.-D. Tan are with the Department of
Electrical and Computer Engineering, University of California at Riverside,
Riverside, CA 92521 USA (e-mail: stan@ee.ucr.edu).}
}
        
\maketitle
\thispagestyle{empty}

\begin{abstract}
  LU factorization for sparse matrices is the most important computing
  step for many engineering and scientific computing problems such as
  circuit simulation. But parallelizing LU factorization with the
  Graphic Processing Units (GPU) still remains a challenging problem
  due to high data dependency and irregular memory accesses. Recently
  GPU-based hybrid right-looking sparse LU solver, called GLU (1.0 and
  2.0), has been proposed to exploit the fine grain level parallelism
  of GPU. However, a new type of data dependency (called double-U
  dependency) introduced by GLU slows down the preprocessing step.
  Furthermore, GLU uses fixed GPU thread allocation strategy, which
  limits the parallelism. In this article, we propose a new GPU-based
  sparse LU factorization method, called {\it GLU3.0}, which solves
  the aforementioned problems. First, it introduces a much more
  efficient data dependency detection algorithm.  Second, we observe
  that the potential parallelism is different as the matrix
  factorization goes on. We then develop three different modes of GPU
  kernel which adapt to different stages to accommodate the computing
  task changes in the factorization.
  %As a result, GLU3.0
  %dynamically allocates GPU blocks and warps based on the
  %number of columns in a level to better balance the computing demands
  %and resources during the LU factorization process.
  Experimental
  results on circuit matrices from University of Florida Sparse Matrix
  Collection (UFL) show that GLU3.0 delivers 2-3 orders of
  magnitude speedup over GLU2.0 for the data dependency
  detection. Furthermore, GLU3.0 achieve 13.0 $\times$ (arithmetic
  mean) or 6.7$\times$ (geometric mean) speedup over GLU2.0 and
  7.1$\times$ (arithmetic mean) or 4.8 $\times$ (geometric mean) over
  the recently proposed enhanced GLU2.0 sparse LU solver on the same
  set of circuit matrices.
\end{abstract}


\section{Introduction}
\label{sec:intro}

Sparse LU factorization plays a critical role in wide engineering and
scientific computing applications such as solving differential and
circuit equations. Particularly, for circuit simulation application
such as the widely used SPICE program~\cite{spice}, the core of the
computing or dominant computing is to solve the linear algebraic
system, $\bm{Ax}=\bm{b}$, resulting from linear or nonlinear circuits
with millions or even billions of extracted components. LU
factorization solves these equations by transforming matrix $A$ into
two matrices: the lower triangular matrix $L$ and upper triangular
matrix $U$ such that $A = LU$. Then the solution $x$ is obtained by
solving the two triangular matrices sequentially. Also for all circuit
simulation problems, matrix $A$ is very large and sparse. As a result,
LU factorization of large sparse matrices becomes a central problem of
those analysis and simulation applications. As VLSI continues to grow
in size, how to improve the LU factorization efficiency and
scalability continues to be a challenging problem.

Graphics processing unit (GPU) provides massive and fine-grain
parallelism with orders of magnitude higher throughput than the CPUs.
For instance, the state-of-the-art NVIDIA Tesla V100 GPU with 5120
cores has a peak performance of over 15~TFLOPS versus about
200--400~GFLOPS of Intel i9 series 8 core
CPUs~\cite{Kirk:Book'13,NvidiaTeslaServer}. Today, in additional to
gaming graphics, GPU has been widely used for more general purpose
computing~\cite{gpgpu} such as EDA, deep learning/AI, finance,
medical and life science etc. However, parallelizing sparse LU
factorization on GPU (GPU) is not straightforward due to high data
dependency and irregular memory access~\cite{HeTan:TVLSI'15}.

% There exists several algorithms targeting sparse matrix LU
% factorization.  SuperLU~\cite{SuperLU} incorporated the idea of
% supernode into Gilbert-Peierls (G/P) left-looking
% algorithm~\cite{Gilbert:SIAM'88}, and there is also a multi-threaded
% version~\cite{SuperLU_MT}. However, the high sparsity of circuit
% matrices makes it hard to form supernodes, thus impeding the
% performance.

There exists some earlier researches targeting parallel sparse LU
factorization on shared memory multi-core CPUs. For instance,
SuperLU\_MT~\cite{SuperLU_MT} is the multi-threaded parallel version
of SuperLU~\cite{SuperLU} for multi-core architectures. However, it
is not easy to form super-node in some sparse matrix such as circuit
matrix. KLU~\cite{Davis:ACMMS'10}, which is specially optimized for
circuit simulation, adopts Block Triangular Form based on
Gilbert-Peierls (G/P) left-looking algorithm~\cite{Gilbert:SIAM'88}
and has become one of the standard algorithms in circuit simulation
applications. The KLU algorithm has been parallelized on multi-core
architecture by exploiting the column-level
parallelism~\cite{Chen:TCASII'11, Chen:TCAD'13}.

Existing GPU based parallel LU factorization solvers mainly focus on
dense matrices,
including~\cite{Galoppo:2005ka,Tomov:2010fk,Tomov:2010jy}. There also
exists a few sparse matrix LU factorization methods on
GPU~\cite{chenhan2011cpu,george2011multifrontal,lucas2010multifrontal}.
But these works mainly convert the sparse matrices into many dense
submatrices (blocks) and then solve them by dense matrix LU
factorizations. However, such strategy may not work well for circuit
matrices, which hardly have dense submatrices.

% UMFPACK~\cite{UMFPACK} is implemented based on a multifrontal
% algorithm. PARDISO~\cite{pardiso_solver} is developed based on a left-right-looking
% algorithm. NICSLU~\cite{Chen:TCAD'13} proposes two parallel schemes, i.e. cluster
% mode and pipeline mode to further accelerate.

Parallel (G/P) left-looking algorithm~\cite{Gilbert:SIAM'88} on GPU
has been explored first in~\cite{RenChen:DAC'12,Chen:TPDS'14}. It
exploits the column-level (called task-level) parallelism due to
sparse nature of the matrix and vector-level parallelism in the
sparser triangular matrix solving in the G/P method. However, the two
loops in triangular matrix solving can't be completely parallelized
(from line 4-8 in Algorithm~\ref{alg:left}) thus the G/P method is
difficult for fine grain parallelization. 

To mitigate this problem, He {\it et. al} proposed a hybrid
right-looking sparse LU factorization on GPU, called GLU
(GLU1.0)~\cite{HeTan:TVLSI'15}. GLU keeps the benefits of the
left-looking method for column-based parallelism and uses the same
symbolic analysis routine. The difference is that it performs the
submatrix update once one column is factorized, which is similar to
the traditional right-looking LU method. However, GLU1.0 used a fixed
scheme to allocate the GPU threads and memory, which limits its
parallelism. Furthermore, the right-looking feature of GLU actually
introduces new data dependency (called double-U dependency in this
paper), which has been reported in GLU2.0~\cite{GLU_UCR} and
\cite{Lee:TVLSI'18}. Double-U dependency can lead to inaccurate
results for some test cases.  Detection of double-U dependency was
added into GLU2.0 to fix this issue, which, however, incurred some
performance degradation compared to GLU1.0. Recently, Lee {\it et al.}
proposed an enhanced GLU2.0 solver~\cite{Lee:TVLSI'18}, which
considers the column count difference in different level, and exploits
some advanced GPU features such as dynamic kernel launch to further
improve the GLU kernel efficiency. However, the fixed GPU threads and
memory allocation method from GLU2.0 for each kernel launch is still
used and limiting performance.

In this article, we propose a new version of GPU-based sparse LU
factorization solver, called {\it GLU3.0}~\footnote{GLU 3.0 source
codes and documents are available at \cite{GLU_UCR}.} for circuit
simulation and more general scientific computing. It is based on
existing GLU1.0/2.0 using hybrid right-looking LU factorization
algorithm. The main improvements of GLU3.0 are summarized as
follows:

\begin{itemize}

\item First, to mitigate the slow process to detect the new double-U
  data dependency in existing GLU2.0 solver, GLU3.0 introduces a new
  dependency detection algorithm. It uses a relaxed principal to find
  all required dependencies, plus some redundant ones. The efficiency
  is a lot higher than the previous solution with little impact on
  performance.

%\item Second, for GLU, in addition to factorization of one column, we
%  observe that a new computing step, called {\it subcolumn update} is
%  also an important measurement of parallel computing task unit for
%  resource allocation. We propose to use subcolumn update as the new
%  parallel computing unit. 

\item Second, we have observed a pattern of potential parallelism as the
    matrix factorization goes on, based on the circuit matrices we
    analyzed. Basically, the number of columns and its associated
    subcolumns (updates) of each column, which are important parallel
    computing task units, are inversely correlated. As a result, we
    can use the number of columns as a good metric for resource
    allocation. We have developed three different modes of GPU kernel that
    adapt to different stages to accommodate the computing task
    changes. As a result, GLU3.0 can dynamically allocate GPU threads
    and memory based on the number of columns in a level to better
    balance the computing demands and resources during the LU
    factorization process.

% \item \textbf{A new method of levelization.} We proposed a new way
% doing levelization. The new method totally fixes the problem of
% additional data dependence with negligible overhead.

% \item \textbf{Optimization of GPU kernels.} We rewrote the GPU kernels
% that run the numerical factorization tasks. As the exploitable
% parallelization opportunities change as the factorization goes on, so
% does the optimal block size and grid size. The new kernels are
% configurable and they always run with an appropriate setting to
% maximize the parallelism.
% Furthermore, the new kernels are resource-aware, which means they can
% take advantage of full computation resources of the GPU to achieve
% shorter runtime.

\end{itemize}

Numerical results on circuit matrices from University of Florida
Sparse Matrix Collection (UFL) show that the GLU3.0 can deliver 2-3
orders of magnitude speedup over GLU2.0 for the data dependency
detection. Furthermore, GLU3.0 {\it consistently} outperforms both GLU
2.0 and the recently proposed enhanced GLU2.0 sparse LU solver on the
same set of circuit matrices. Furthermore, GLU3.0 achieve 13.0
$\times$ (arithmetic mean) or 6.7 $\times$ (geometric mean) speedup
over GLU 2.0 and 7.1 $\times$ (arithmetic mean) or 4.8 $\times$
(geometric mean) over recent proposed enhanced GLU2.0 sparse LU solver
on the same set of circuit matrices.

This article is organized as follows. Section~\ref{sec:review} gives a
brief review of previous work on sparse matrix LU factorization, GPU
programming, and GLU itself. In Section~\ref{sec:new_method}, we
present our two novel improvements. Several numerical examples
and related discussions are presented in Section~\ref{sec:results}. At
last, Section~\ref{sec:conclusion} concludes this work.

\section{Review of LU factorization and CUDA}
\label{sec:review}
In this section, we briefly review the traditional G/P left-looking
method for sparse matrices LU factorization~\cite{Gilbert:SIAM'88} and
the recently proposed hybrid right-looking algorithm used in GLU1.0,
GLU2.0 and a recent GLU enhancement work~\cite{Lee:TVLSI'18}. We would
also briefly review the GPU architectures and NVIDIA CUDA programming
system.

\begin{figure}
    \centering
    \includegraphics[width=0.5\columnwidth]{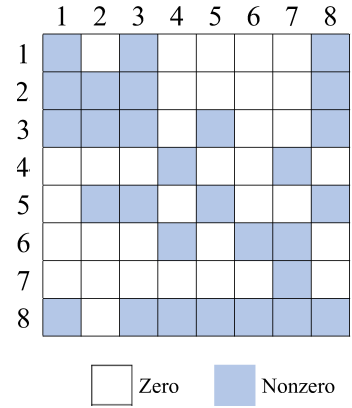}
    \caption{The example matrix}
    \label{fig:example_mat}
\end{figure}

An example matrix is used for illustrating important concepts and
algorithms in the following discussions. The matrix is shown in
Fig.~\ref{fig:example_mat}, where the colored spots stand for nonzero
elements.

%Typically, circuit simulations start with the nonlinear differential
%modified nodal analysis (MNA) and form the circuit equations
%from VLSI circuits~\cite{Vlach:Book'94}. \rev{For transient analysis, the
%differential MNA equations will first be linearized by Newton-Raphson
%(NR) iteration in the first loop. Then, in the second loop, the
%equations will be time discretized into algebraic linear equation in
%the form of $Ax = b$, which will then be solved by LU factorization.}
%LU factorization of a $n \times n$ matrix $A$ has $O(n^s)$ time
%complexity, where $n$ is the size of the circuit matrix and exponent
%$s$ is about 1.1 to 1.5 in general.

\subsection{Left-looking factorization method}
Traditional Gaussian elimination based LU factorization method (also
called right-looking method) solves one row for $U$ matrix and then
one column for $L$ matrix in each iteration. While the G/P
left-looking method computes one column in one iteration for both $L$
and $U$ instead, which is achieved by solving a lower triangular
matrix. It also allows the symbolic fill-in analysis of $L$ and $U$
matrices before the actual numerical computing. As a result, G/P
left-looking method shows better performance for sparse matrices,
especially those from circuit simulations~\cite{Davis:Book'06}. 

Algorithm~\ref{alg:left} shows the detailed implementation of G/P
left-looking LU factorization~\cite{Gilbert:SIAM'88}. The input of
this algorithm $A_s$ is the nonzero filled-in matrix of $A$ after
symbolic analysis. 
% Preprocessing includes fill-in, permutation
% and scaling, which due to the paper length limit would not be
% discussed in this paper.
The matrix $A_s$ is factorized column by column (the outer $j$
loop), and factorizing each column for both $L$ and $U$ contains two
steps.
The first step (lines 4-9) is to solve a triangular matrix. In each
$k$ loop, element-wise multiply-and-accumulate (MAC) operation is done
(line 6-8) for the partial column vector $A_s(k+1:n,j)$. $A_s(i,k)$
are the elements in the factorized columns on the left of current
column $j$. This is the reason why it is
named left-looking LU method. Then the second step (lines 10-13) is a
much simpler loop that finishes the factorization of this column.
Triangular matrix solving (lines 4-9) is
the most essential and computationally expensive step in this
algorithm.

\begin{figure}
    \centering
    \includegraphics[width=0.8\columnwidth]{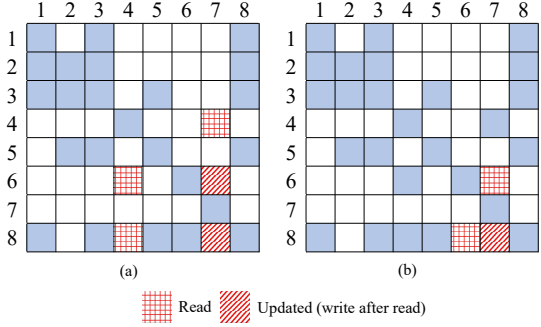}
    \caption{The two update iterations completing
    factorization of the $7$th column ($j=7)$ (a) update using the 4th
    column ($k=4$) (b) update using the 6th column ($k=6$)}
    \label{fig:mat_update}
\end{figure}

Fig.~\ref{fig:mat_update} gives a complete example of this step. In
this example, column 7 is being factorized, meaning $j=7$ in
Algorithm~\ref{alg:left}. Only two $k$'s satisfy $A_s(k,j)\neq0$ (line
4), which are $4$ and $6$ (as $A_s(4,7) \neq0$ and $A_s(6,7)\neq0$).
The two sub-figures show these two iterations respectively. In (a),
$k=4$, so column 4 is used to update column 7. The update operation
refers to lines 6-8 of Algorithm~\ref{alg:left}, where two elements of
column 7 ($A_s(6,7) $ and $A_s(8,7)$) are updated by MAC operations
with the red elements in column 4 multiplying $A_s(4,7)$. (b) shows
the next iteration, where $k=6$, column 3 is used to further update
column 7, which can be explicitly written as $A_s(8,7)\leftarrow
A_s(8,7) - A_s(8,6) * A_s(6,7)$.

\begin{algorithm}[ht]
  \caption{The Gilbert-Peierls left-looking algorithm}
\label{alg:left}
\begin{algorithmic}[1]
\STATE /* Scan each column from left to right */
\FOR{$j=1$ to $n$}
\STATE /*Triangular matrix solving */
\FOR{$k=1$ to $j-1$ where $A_s(k,j)\not=0$}
\STATE /*Vector multiple-and-add (MAC) */
\FOR{$i=k+1$ to $n$ where $A_s(i,k)\not=0$}
\STATE $A_s(i,j)=A_s(i,j)-A_s(i,k)*A_s(k,j)$
\ENDFOR
\ENDFOR
\STATE /*Compute column j for L matrix*/
\FOR{$i=j+1$ to $n$ where $A_s(i,j)\not=0$}
\STATE $A_s(i,j)=A_s(i,j)/A_s(j,j)$
\ENDFOR
\ENDFOR
\end{algorithmic}
\end{algorithm}

\subsection{Review of the column-based right-looking algorithm used in
GLU}
\label{subsec:col_right_look}
As elaborated in \cite{HeTan:TVLSI'15}, the G/P left-looking sparse LU
factorization has one limitation that it failed to parallelize the two
loops in triangular matrix solving process (lines 4-8 of
Algorithm~\ref{alg:left}). It can only work on (write) one column
(current column $j$) at a time as indicated in line 7. To mitigate
this problem, He {\it et al.} proposed the hybrid column-based
right-looking LU factorization algorithm for
GLU~\cite{HeTan:TVLSI'15}. The algorithm is hybrid because it still
keeps the column-based parallelism in the left-looking algorithm
while updates columns on the right during factorization. Similar
symbolic analysis is still applied in advance as well.

\begin{figure}
    \centering
    \includegraphics[width=0.6\columnwidth]{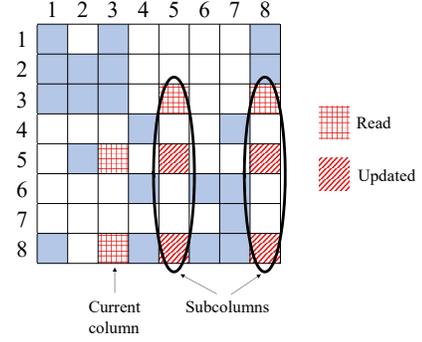}
    \caption{Subcolumns and submatrix column 3. All highlighted
        elements compose the submatrix, which include elements being
        read and elements being updated.}
    \label{fig:submat}
\end{figure}

%$A_s(x,y)$
%stands for the LU symbolically factorized $A$ matrix, where memory is
%allocated for all the fill-ins and non-zero elements, and values of
%non-zero elements have been copied. The final result after
%factorization is $A_s = L + U - I$, where $I$ is identity matrix.
The hybrid right-looking LU factorization algorithm is listed in
Algorithm~\ref{alg:right_look}. Similarly, the current column under
computing
is indexed by $j$. For each column, the first step is to compute the
$L$ part of the current column (lines 4-6), which is equivalent to
lines 10-12 of Algorithm~\ref{alg:left}. Then, it looks right
to find all columns $k$ ($k > j$) that meet $A_s(k,j)\not=0$, and
uses the currently factorized column $j$ to update these columns
(lines 8-12).  
For the sake of presentation convenience without confusion, we name
these columns \textit{subcolumns} of column $j$. Note that these
subcolumns are not part of the column $j$. Furthermore, this step of
updating all subcolumns is called \textit{submatrix update}, where all
elements being read or updated form a \textit{submatrix}.
Fig.~\ref{fig:submat} gives an example illustrating these two
concepts. In the figure, $j=3$, its subcolumns are column 5 and 8,
because $A_s(3,5)$ and $A_s(3,8)$ are nonzero elements. Corresponding
this to the execution of Algorithm~\ref{alg:right_look}, during
iteration $j=3$, two $k$'s meet the condition of line 8, which are 5
and 8.

The key difference between this right-looking algorithm and the
left-looking one is that submatrix update completes the equivalent
jobs of triangular matrix solving (lines 4-9 of
Algorithm~\ref{alg:left}) in advance. In the example shown in
Fig.~\ref{fig:mat_update}, both update operations are completed while
$j=7$. However, in the case of the right-looking algorithm, the update
in (a) is done while $j=4$, and update in (b) is done while $j=6$. As
will be discussed in detail in the following section, this difference
enables exploiting parallelization between subcolumns.

\begin{algorithm}[htb]
  \caption{The hybrid column-based right-looking algorithm for GLU1.0/2.0}
\label{alg:right_look}
\begin{algorithmic}[1]
\STATE /* Scan each column from left to right */
\FOR{$j=1$ to $n$}
\STATE /*Compute  column $j$ of L matrix*/
\FOR{$k=j+1$ to $n$ where $A_s(k,j)\not=0$}
\STATE $A_s(k,j)=A_s(k,j)/A_s(j,j)$
\ENDFOR
\STATE /*Update the submatrix for next iteration*/
\FOR{$k=j+1$ to $n$ where $A_s(j,k)\not=0$}
\FOR{$i=j+1$ to $n$ where $A_s(i,j)\not=0$}
\STATE $A_s(i,k)=A_s(i,k)-A_s(i,j)*A_s(j,k)$
\ENDFOR
\ENDFOR
\ENDFOR
\end{algorithmic}
\end{algorithm}

\subsection{Additional data dependency in GLU: the fix in GLU2.0}
\label{sec:GLU2.0}
Data dependency is an important issue in parallel computing or
general high performance computing. It puts hard requirements in the
orders of operations. In SuperLU~\cite{SuperLU_MT} and
NICSLU~\cite{Chen:TCAD'13}, elimination tree has been used to resolve
this issue. Similarly, methods like dynamic application dependency
graph is used in heterogeneous computing~\cite{Xiao:ICCAD'17}. For
GLU, in order to factorize several columns in parallel, data
dependency
between columns needs to be detected in the first place. With complete
information of dependency, columns can be grouped into
\textit{levels}, where all columns in the same level are independent
of each other and can thus be factorized in parallel. Such process
deriving information about levels is called \textit{levelization},
which is a similar method to elimination tree. In the
left-looking LU factorization method, levelization is done by studying
the sparsity pattern of the $U$ matrix. Any $U(i, j) \neq 0, i < j$
results in column $j$ being dependent on $i$ because of the triangular
matrix solving (lines 4-9 in Algorithm~\ref{alg:left}). This
dependency detection algorithm was also used in GLU1.0.

\begin{figure}
    \centering
    \includegraphics[width=0.8\columnwidth]{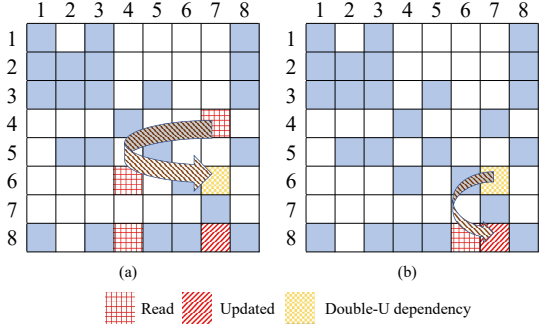}
    \caption{An example of double-U dependency originated from
    element (6,7)}
    \label{fig:double_u_dep}
\end{figure}

%Algorithm~\ref{alg:dependency_old} lists the complete flow of this.
%
%\begin{algorithm}[htb]
%\caption{Column dependency detection algorithm used in GLU1.0}
%\label{alg:dependency_old}
%\begin{algorithmic}[1]
%\FOR{$i=1$ to $n$}
%    \STATE /* Look up for all nonzeros in column $i$ of $U$ */
%    \FOR{$j=1$ to $i-1$ where $A_s(j,i)\not=0$}
%        \IF{Column $j$ of $L$ is not empty}
%            \STATE Add $j$ to $i$'s dependency list
%        \ENDIF
%    \ENDFOR
%\ENDFOR
%\end{algorithmic}
%\end{algorithm}
However, as reported in \cite{GLU_UCR, Lee:TVLSI'18}, the hybrid
right-looking algorithm used in GLU leads to a new column dependency
named \textit{double-U dependency}, originating from the read-write hazard
during parallel submatrix updates. An example of this can be found in
columns 4 and 6 of the example matrix, with the details highlighted in
Fig.~\ref{fig:double_u_dep}. In (a), $A_s(6,7)$ is updated by column
4: $A_s(6,7) \leftarrow A_s(6,7) - A_s(6,4) * A_s(4,7)$. In (b),
$A_s(6,7)$ is used to update column 7: $A_s(8,7) \leftarrow A_s(8,7) -
A_s(8,6) * A_s(6,7)$. In the scheme of GLU1.0, both updates are
executed in parallel. However, $A_s(6,7)$ is written in (a) and read
in (b), which forms a read-write hazard when they are executed
in parallel. To ensure correctness, the write operation in (a) must
finish before the read operation in (b). As a result, an additional
dependency between columns 4 and 6 needs to be introduced
undesirably.

Such read-write dependency is called double-U dependency in
GLU2.0 as it originates from two overlapped U-shaped dependencies as
shown in Fig.~\ref{fig:double_u_dep}. To detect this new dependency,
GLU2.0 introduced a different dependency detection process as shown in
Algorithm~\ref{alg:dependency_2u}. This algorithm directly
looks for double-U dependency. Suppose $k$ is found for given $i, t$
and $j$, $A_s(t, k)$ is updated by $A_s(t, i)$, while it is also used
to update $A_s(j, k)$. As a result a double-U dependency exists
between columns $i$ and $t$. In the example of
Fig.~\ref{fig:double_u_dep}, $i = 4, t = 6, j = 8$, and $k = 7$
respectively.

\begin{algorithm}[htb]
\caption{Double-U dependency detection algorithm used in GLU2.0}
\label{alg:dependency_2u}
\begin{algorithmic}[1]
\FOR{$i=1$ to $n$}
    \STATE Store all non-zero indices of row $i$ in $I_i$
    \FOR{$t=i$ to $n$ where $A_s(t,i) \neq 0$}
        \FOR{$j=t$ to $n$ where $A_s(j,t) \neq 0$}
            \STATE Store all non-zero indices of row $j$ in $I_j$
            \IF{$\exists k, k \in I_i, k \in I_j, k > t$}
                \STATE Add $i$ to $t$'s dependency list
            \ENDIF
        \ENDFOR
    \ENDFOR
\ENDFOR
\end{algorithmic}
\end{algorithm}

However, this detection algorithm can be quite expensive because of
the three nested loops that have $O(n^3)$ complexity. In comparison,
there are only two for loops in the $U$ matrix pattern based
dependency detection algorithm. It leads to performance degradation
compared to GLU1.0.

\begin{figure}[htb]
    \centering
    \includegraphics[width=0.8\columnwidth]{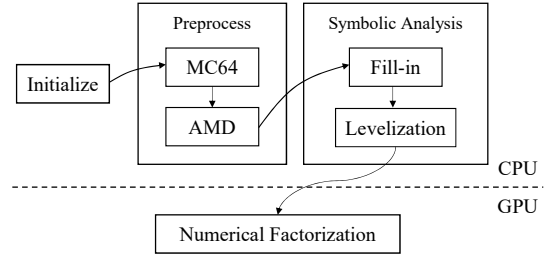}
    \caption{Complete flow of GLU2.0}
    \label{fig:flow}
\end{figure}

Besides dependency detection and levelization, some preprocessing and
symbolic analysis needs to be done on CPU ahead of factorization. The
preprocessing includes MC64 and AMD (Approximate minimum degree)
algorithms in order to reduce the number of final nonzero elements, as
is done in NICSLU~\cite{Chen:TCAD'13}. Symbolic analysis includes
fill-in and levelization. Combining all this, we have the complete
flow of GLU2.0 shown in Fig.~\ref{fig:flow}.

\subsection{Enhancements to GLU2.0}
Recently, Lee \textit{et. al} proposed a method to enhance
GLU2.0\cite{Lee:TVLSI'18}. In detail, three different kernels were
proposed, namely cluster mode, batch mode and pipeline mode. Modes are
selected based on different number of columns in levels. In batch
mode and pipeline mode, overlapped execution of different levels is
achieved to some extent, which contributes to the speed-up. Besides
this, kernel launches are managed by a small kernel instead of CPU,
which is called dynamic parallelism enabled by CUDA compute capability
3.0. Combining these techniques, the enhanced GLU has achieved
1.26$\times$ (geometric mean) speedup over GLU2.0.

%Cluster mode works in the same way as the kernel of previous GLU
%versions and are good for large levels. Batch mode is designed for
%levels with two columns and pipeline mode is for levels with one
%column. In both modes, kernels for multiple levels are launched
%\textit{in batches}. Base on the observation that columns in
%proceeding levels are typically dependent on only part of the previous
%level, the factorization of them can start as long as the 
%dependent columns have finished updating their subcolumns. In this
%way, execution of different levels can be overlapped to some extent.
%A small kernel is used to launch child kernels, using dynamic
%parallelism enabled by CUDA compute capability 3.0. However, dynamic
%parallelism is usually used where recursive kernel launches are
%required, and where more refined computation is needed. It does not
%help much if it used just to manage the computing resources.

%Recently, an enhanced GLU solver was proposed in~\cite{Lee:TVLSI'18},
%in which the authors proposed to use dynamic kernal launch features of
%new GPU to avoid more kernal lanuches from CPU first.  Also to explore
%the facts that many levels have just a few or just one column, so the
%kernal launches can be optimizee by using different executation models
%for diffferent levels such as cluster mode for levels with many
%columns, batch model for level with a few column and pipeline for
%level with only one column. By combined those techniques, the enhanced
%GLU deliver 2.49 (geometric mean) speedup over GLU 2.0 over some
%circuit matrices. 

\subsection{Review of GPU Architecture and CUDA programming}
\label{subsec:gpu_review}
\begin{figure}[tb]
  \centering
  \includegraphics[width=1.0\columnwidth]{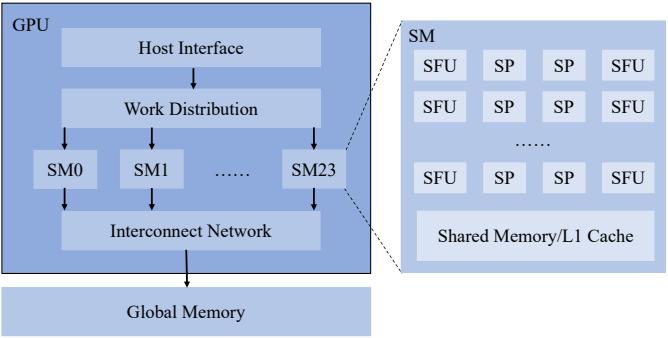}
   \caption{Diagram of NVIDIA TITAN X and the streaming multiprocessor.
   (SP is short for streaming processor, L/S for load/store unit,
   and SFU for Special Function Unit.)}
   \label{fig:gpu}
\end{figure}

CUDA, short for Compute Unified Device Architecture, is
the parallel programming model for NVIDIA's general-purpose GPUs. The
architecture of a typical CUDA-capable GPU is consisted of an array of
highly threaded streaming multiprocessors (SM) and comes with a
huge amount of DRAM, referred to as global memory. Take the GTX
TITAN X GPU for example. It contains 24 SMs, each of which has 128
streaming multiprocessors (SPs, or CUDA cores called by NVIDIA), 8
special function units (SFU), and its own shared memory/L1 cache. The
architecture of the GPU and streaming multiprocessors is shown in
Fig.~\ref{fig:gpu}.

\begin{figure}[tb]
  \centering
  \includegraphics[width=1.0\columnwidth]{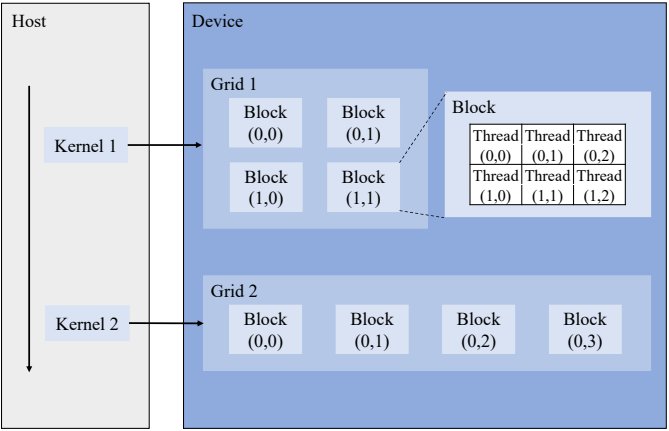}
  \caption{The programming model of CUDA.}
  \label{fig:CUDA_model}
\end{figure}

As the programming model of GPU, CUDA extends C into CUDA C and
supports such tasks as threads calling and memory allocation, which
makes programmers able to explore most of the capabilities of GPU
parallelism. In CUDA programming model, illustrated in
Fig.~\ref{fig:CUDA_model}, threads are organized into blocks; blocks
of threads are organized as grids. CUDA also assumes that both the
host (CPU) and the device (GPU) maintain their own separate memory
spaces, which are referred to as host memory and device memory
respectively. For every block of threads, a shared memory is
accessible to all threads in that same block. The global memory is
accessible to all threads in all blocks. Developers can write programs
running millions of threads with thousands of blocks in parallel. This
massive parallelism forms the reason that programs with GPU
acceleration can be much faster than their CPU counterparts.  CUDA C
provides its extended keywords and built-in variables, such as
\texttt{blockIdx.\{x,y,z\}} and \texttt{threadIdx.\{x,y.z\}}, to
assign unique ID to all blocks and threads in the whole grid
partition. Therefore, programmers can easily map the data partition
to the parallel threads, and instruct the specific thread to compute
its own responsible data elements. Fig.~\ref{fig:CUDA_model} shows an
example of 2-dim blocks and 2-dim threads in a grid, the block ID and
thread ID are indicated by their row and column positions.

\section{New GPU sparse LU solver: GLU 3.0}
\label{sec:new_method}
As introduced above, the work flow of factorizing a sparse matrix with
GLU can be divided into two parts: the preprocessing and symbolic
analysis on CPU and the numeric factorization on GPU. The second part
on GPU might be repeated for many times when solving a nonlinear
equation with Newton-Raphson method in circuit simulation. In this
work, we significantly improve both the symbolic analysis and numeric
factorization.

\subsection{Relaxed data dependency detection method for GLU}
\label{sec:dep_alg}
%Levelization is a critical step for parallel computation. Due to
%dependency between levels, dependency analysis and scheduling have to
%be done prior to numerical factorization, which is similar to the
%idea of Elimination Tree in \cite{nicslu, superlu}. Basically,
%levelization aims to find out groups of columns that can be factorized in
%parallel by studying the dependency between columns. Once all
%dependency is detected, BFS can be used to group columns into levels.
%So the key to levelization is dependency detection.

As mentioned in Section~\ref{sec:GLU2.0}, the prior dependency
detection algorithm introduced to cover double-U dependency slowed
down the factorization a lot. In this work, we solve this problem by
proposing a better dependency detection algorithm, called {\it
  relaxed} column dependency detection method, which can reduce the
process down to two loops. The new algorithm is based on the
observation that a {\it necessary condition} for such additional
dependency is the existence of nonzero elements on the left of
diagonal element in the $L$ matrix. In the example in
Fig.~\ref{fig:double_u_dep}, such dependency exists between columns 4
and 6. The nonzero element $A_s(6,4)$ on the left of diagonal element
$A_s(6,6)$ is the necessary condition that column 6 depending on
column 4, as it is the reason that $A_s(6,7)$ gets updated, and
$A_s(6,7)$ is the very element that induces the double-U dependency.

\begin{figure}[h]
    \centering
    \includegraphics[width=0.7\columnwidth]{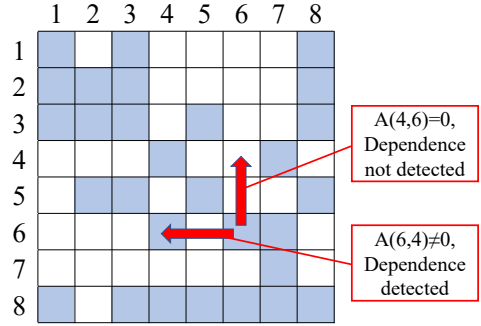}
    \caption{Comparison of left looking and up looking, left looking
    is able to detect double-U dependency.}
    \label{fig:newdetect}
\end{figure}

Based on this observation, the new method simply just look 
for nonzero elements on the left of diagonal element, which can be
called simply as ``left looking'', to find such new dependency. It is
very similar to the ``up looking'' in the $U$ matrix based dependency
detection method employed in the left-looking factorization algorithm.
Fig.~\ref{fig:newdetect} compares the result of them by applying both
methods to column 6. As there is no nonzero element in column 6 of $U$
matrix, ``Looking up'' from $A_s(6,6)$ will find no depended column of
column 6. On the other hand, ``looking left'' from the same element, a
nonzero element in column 4 can be seen, which is interpreted as the
new dependency between columns 4 and 6 that is the double-U
dependency as expected. The complete algorithm incorporating the new
dependency detection method is listed in
Algorithm~\ref{alg:dependency}. Lines 8-11 are the additional ``left
looking'' part that is added to the original dependency detection
algorithm from GLU1.0.

\begin{figure}
    \centering
    \includegraphics[width=1.0\columnwidth]{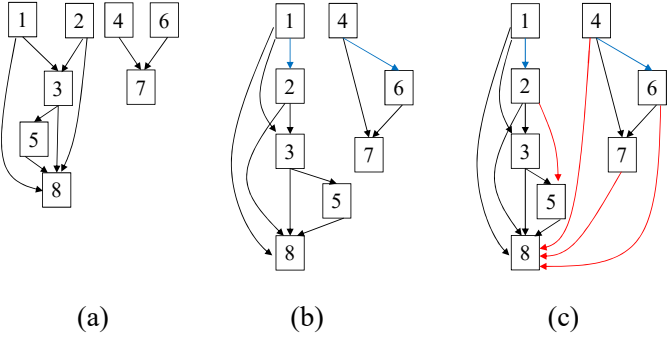}
    \caption{Dependency graph generated from 3 methods: (a) GLU1.0:
    incorrect result (b) GLU2.0: correct result (c) This work:
    the relaxed data dependency}
    \label{fig:3level}
\end{figure}

In order to compare the aforementioned three dependency detection
methods, they are applied to the example matrix from
Fig.~\ref{fig:mat_update} and the results are shown in
Fig.~\ref{fig:3level} respectively. An edge $x\rightarrow y$ indicates
that that column $x$ depends on column $y$. Comparing (a) and (b), the
extra dependencies $1\rightarrow 2$ and $4\rightarrow 6$ (marked
by blue line) are the double-U dependencies. Further comparing (b) and
(c), we can see that the proposed method is able to detect all
required column
dependencies, plus a few redundant ones marked by red. Despite the
redundant dependencies, the result of levelization is exactly the
same, which means the same numerical performance on GPU can be
expected. This example shows that the redundant dependencies have
minor, if none, impacts on parallelism exploration of GLU. The reason
why this dependency detection method is called {\it relaxed} is that
it does not detect the exact set of dependencies, but a sufficient one
possibly with some redundant dependencies. More examples about this
will be reported later in Section~\ref{sec:results}.

\begin{algorithm}[htb]
\caption{The proposed relaxed column dependency detection method}
\label{alg:dependency}
\begin{algorithmic}[1]
\FOR{$k=1$ to $n$}
    \STATE /* Look up for all nonzeros in column $k$ of $U$ */
    \FOR{$i=1$ to $k-1$ where $A_s(i,k)\not=0$}
        \IF{Column $i$ of $L$ is not empty}
            \STATE Add $i$ to $k$'s dependency list
        \ENDIF
    \ENDFOR
    \STATE /* Look left for all nonzeros in row $k$ of $L$ */
    \FOR{$i=1$ to $k-1$ where $A_s(k,i)\not=0$}
        \STATE Add $i$ to $k$'s dependency list
    \ENDFOR
\ENDFOR
\end{algorithmic}
\end{algorithm}

\subsection{New numerical kernels}
\label{sec:newkernel}

Before we discuss our new GPU kernels, we would like to first review
the submatrix update in GLU, which is a key step
in Algorithm~\ref{alg:right_look}.

\subsubsection{The submatrix update revisited}

The submatrix update is explicitly listed in
Algorithm~\ref{alg:submatrix_update} below. 
\begin{algorithm}[htb]
  \caption{The submatrix update in GLU}
\label{alg:submatrix_update}
\begin{algorithmic}[1]
\STATE /*Update the submatrix for next iteration*/
\FOR{$k=j+1$ to $n$ where $A_s(j,k)\not=0$}
\FOR{$i=j+1$ to $n$ where $A_s(i,j)\not=0$}
\STATE $A_s(i,k)=A_s(i,k)-A_s(i,j)*A_s(j,k)$
\ENDFOR
\ENDFOR
\end{algorithmic}
\end{algorithm}
Specifically, we can write the submatrix to be updated as 
\begin{equation}
A_{sub} =
\left[ {\begin{array}{ccc}
  A_s(j+1, j+1) & \cdots &  A_s(j+1,n) \\
   \vdots & \ddots & \vdots \\
  A_s(n, j+1) & \cdots & A_s(n,n)  \\
 \end{array} } \right]  \\
\end{equation}
where the size of the submatrix is $N \times N$, with $N = n-j$. The
submatrix update operation can be further represented in the following
format:
\begin{align}
\label{eq:submatrix_update}
  A_{sub}  & \leftarrow   A_{sub} \nonumber \\
  & - \left[ {\begin{array}{c}
  A_s(j+1, j)  \\
  \vdots \\
  A_s(n, j) \\
 \end{array} } \right] \cdot \left[A_s(j,j+1), \cdots, A_s(j,n) \right]
\end{align}
where the size of the two vectors are $N \times 1$, and $1 \times N$.
Both two vectors and $A_{sub}$ matrix
are sparse. From~\eqref{eq:submatrix_update}, we can see that the
submatrix update consists of two operations: (a) vector tensor
multiplication (the second item on the right hand side); (b) matrix
addition.
%As a result, we can see that the two steps can be easily
%parallelized.

In the implementation of GLU, the submatrix update 
\begin{align*}
&A_{sub} - \left[ {\begin{array}{c}
   A_s(j+1, j)  \\
   \vdots \\
   A_s(n, j) \\
  \end{array} } \right] \cdot \left[A_s(j,j+1), \cdots,
  A_s(j,n) \right]  \nonumber \\ 
\end{align*}
is done in a
column-wise way as depicted in \eqref{eq:subcolumn_update}:
\begin{align}
\label{eq:subcolumn_update}
  & \vec{A_s}(j+1:n, i) - \vec{A_s}(j+1:n, j)\cdot A_s(j,i),
     \nonumber \\
  & \mbox{for} \,\, i = [j+1, \cdots, n]
\end{align}
where
\begin{align*}
  \vec{A_s}(j+1:n, i)   & = [A_s(j+1, i) , \dots,  A_s(n, i)   ]^T \\
  \vec{A_s}(j+1:n, j) & = [A_s(j+1, j) , \dots, A_s(n, j) ]^T. 
\end{align*}

As we can see, the submatrix update consists of vector operations
or {\it subcolumn update}. Each time, we can update one {\it subcolumn
}$i$ as shown in \eqref{eq:subcolumn_update}. This can be parallelized in
GPU where each resulting element can be computed using one thread, where
the operation is multiply-accumulate (MAC) operation. There are two
levels of parallelism: namely (a) the vector operations (or subcolumn
updates) for different vectors as shown in \eqref{eq:subcolumn_update}
and (b) element-wise MAC operations in each vector or subcolumn. In
contrast, the left-looking algorithm only has element-wise MAC operation
parallelism in the triangular matrix solving process.

\subsubsection{New adaptive GPU kernel}
\label{sec:new_kernel}

The second contribution we made is to significantly improve the GPU
kernel computing efficiency for GLU. GLU1.0/2.0 used fixed resource
allocation strategy in the GPU kernel. However, as the matrix
size grows, the fixed resource allocation strategy will significantly
restrict the potential parallelism in GPU.

Before going into details, we define several terms for the
ease of discussion. All columns in the same level that can be
factorized in parallel referred to as \textit{parallelizable}; the
\textit{size} of a level is the number of parallelizable columns in
this level. In other words, a \textit{large} level has many
parallelizable columns, while a \textit{small} level has few columns.

As defined, all columns in one level are parallelizable, and
each column has many associated parallelizable subcolumns. This
two-level parallelism distinguishes GLU from other parallel sparse
matrix LU factorization algorithms. Two metrics can be used to
describe the potential parallelism respectively, namely the size
of one level, and the maximum number of subcolumns for all columns in
one level, because they are the basic units that get parallelized.
%The first one is
%self-explanatory, and the second one is based on the fact that
%updating the submatrix with more subcolumns generally involves more
%calculations.

\begin{figure}[h]
    \centering
    \begin{subfigure}[Level versus its size and the maximum number of
      subcolumns]{\includegraphics[width=0.9\columnwidth]{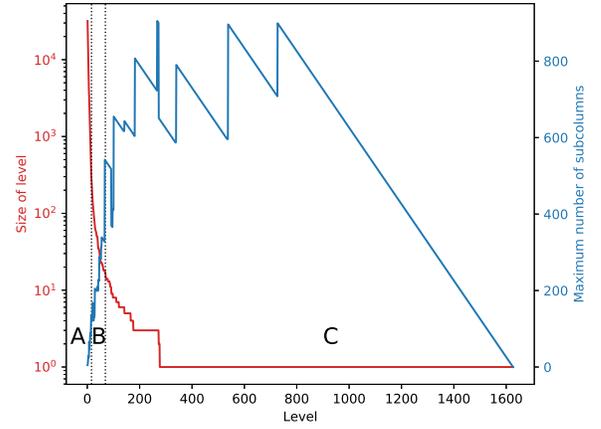}}
    \end{subfigure}
    % \quad
    \\
    \hspace{0.1in}
    \begin{subfigure}[Zoomed in view]{\includegraphics[width=0.9\columnwidth]{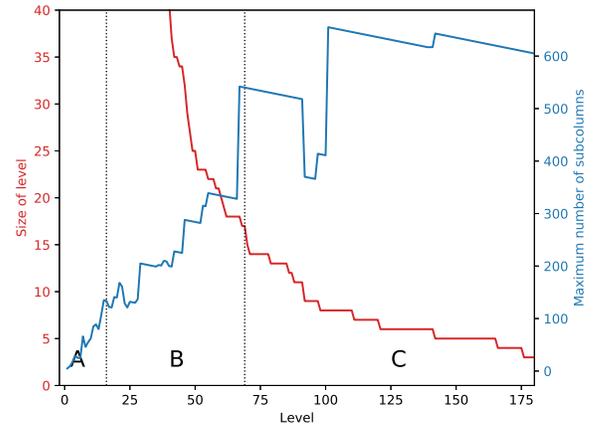}}
    \end{subfigure}
    \caption{Number of columns and subcolumns of different levels. Maximum
    number of subcolumns is used for each level. The matrix is ASIC100ks
    from \cite{Davis:UFS'11}.}
    \label{fig:levels}
\end{figure}

The potential parallelism keeps changing
across the levels, which is the key reason of the fixed resource
allocation strategy being inefficient. The trend of potential
parallelism is shown in Fig.~\ref{fig:levels}. An important
observation is that {\it levels generally fall into three categories},
which are also labeled in the figure. Type A levels in the beginning
stage of factorization have huge number of parallelizable columns,
while each column has very few associated subcolumns. For higher
throughput, parallelizing columns should be prioritized for this type
of levels. Type C levels, in contrast, have limited number of columns,
while each column generally has large number of subcolumns until very
end of the factorization process. As a result parallelizing subcolumns
is more important for this type of levels. Type B levels, in the
transitional stage, have great numbers of columns, and at the same
time columns also have many subcolumns. So parallelism should be
naturally balanced between them.

Furthermore, the second important observation we have is rev{{\it that the
  in one level number of parallelizable columns and their associated subcolumns are
  inversely correlated in general}. As a result, we can use the size
  of a level as a good estimation of the associated subcolumn
numbers to dynamically allocate the computing resources to further
improve the GPU kernel computing efficiency. Based on this
observation, we propose three computing modes of GLU kernels, which
are chosen based on the level sizes in a progressive way to
accommodate the three types of levels.}

\begin{figure}[t]
    \centering
    \includegraphics[width=\columnwidth]{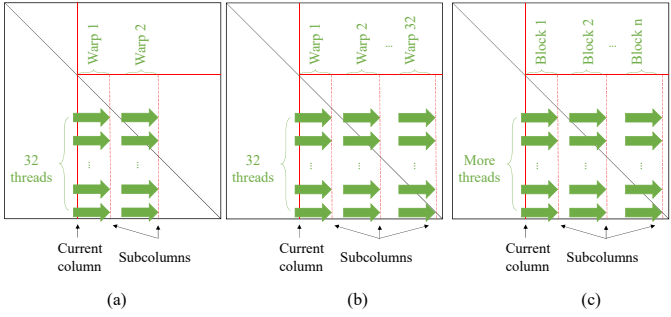}
    \caption{Comparison of the concurrency layout for one column
        in different kernels: (a) Small block mode (b) Large block
        mode (c) Stream mode}
    \label{fig:layout}
\end{figure}

\begin{numberlist}

\item {\bf Small block mode:} This mode is designed for type A levels.
  A convention we have followed from GLU for this mode is that one
  block takes care of a column, and one warp is assigned to a
  subcolumn. In this mode, as shown in
  Fig.~\ref{fig:layout}(a), fewer warps are assigned to a CUDA block,
  which is why we name it small block mode. As the total number
  of warps is fixed for a given GPU, more blocks, or equivalently the
  factorization of more columns, can be carried out in parallel, which
  fits the requirement of type A levels: huge number of columns with a
  few subcolumns. Another important observation from
  Fig.~\ref{fig:levels} is that the number of subcolumns is gradually
  increasing, and the level size is decreasing quickly. In order to
  adapt to this change, the number of warps assigned to a block is
  gradually increased, assisting the growing number of subcolumns
  and trying to make full use of available warps at the
  same time. The number of warps assigned to a block grows from 2 to
  4, 8, and eventually to 32, which is the number in the next
  mode. The exact number of warps assigned to a block is determined by
  number of columns in a level using following expression:
  \begin{equation}
      W  = \frac{\text{Total number of warps}}{\text{Level size}}
  \end{equation}
  where $W$ is the number of warps assigned to each block.

  Another factor limiting the number of possible parallel columns is
  memory. Because the columns being factorized are stored as a dense
  form in global memory, too many columns from a big level can
  overflow the memory. Specifically, during factorization of each
  column, an array of size $n$ is allocated for caching. As a result,
  the maximum parallelizable columns $N$ can be calculated as:
  \begin{equation}
    \label{eq:memlimit}
      N  = \frac{\text{Max global
      memory allowed}}{n * \text{sizeof(float)}}
  \end{equation}
  where $n$ is the number of rows of the matrix.
    
\item {\bf Large block mode:} This mode takes care of type B levels,
  and it is similar to the kernel used in previous GLU versions. Same
  as the small block mode, each block takes care of one column and
  each warp is assigned to a subcolumn. In this
  mode, the number of subcolumns still keep growing, the number of
  threads each subcolumn gets (32, one warp) becomes
  insufficient. However, the maximum number of a thread block (1024)
  prohibits any further increase in this number.

\item {\bf Stream mode:} To tackle maximum warp size (32) problem,
  Stream mode is proposed for type C levels in this work. In this
  mode, blocks instead of warps are assigned to each subcolumn, and
  therefore kernel calls instead of blocks are assigned to each
  column, and a block is assigned to each subcolumn now, as is
  shown in Fig.~\ref{fig:layout}(c). To fully exploit
  parallelism within the same level, CUDAStreams are used, which
  allows parallel kernel execution through streams in a GPU. Although
  the number of CUDAStreams could also be dynamically, it has been
  observed that creating
  more CUDAStreams sometimes has a negative effect in performance. As
  a result, the number of CUDAStreams has been set to a fixed number,
  16. This number is able to produce optimal results based on our
  experimental results which will be discussed later.
  Accordingly, this mode begins as long as the level size drops to 16.

\end{numberlist}

We remark that the three GPU kernel modes we proposed are quite
different than then three modes proposed in~\cite{Lee:TVLSI'18}. First,
our approach is based on the observation of both parallelizable column count and
associated subcolumn count change with different levels, while Lee's
work is only based on the column count in each level. Second, we
propose to dynamically allocate GPU computing resources (different
number of warps and threads in each blocks etc) based on those
information, while Lee's work exploits some advanced GPU features such
dynamic kernel launch. Third, Lee's work focus more on exploit
parallelism between different levels, while GLU 3.0 focus on
dynamically changing parallelisms in one level over the course of
factorization process.

\section{Numerical results and discussions}
\label{sec:results}

\begin{table*}[t]
\centering
\caption{Solver runtimes of GLU3.0 vs previous works, where {\it
        nz } stands for number of nonzeros before fill-in, and {\it
        nnz} stands for number of nonzeros after fill-in}
\label{tbl:kernel}
\begin{tabular}{|l|l|l|l|l|l|l|l|l|l|l|l|l|}
\hline
\multirow{3}{*}{Matrix}    & \multirow{3}{*}{\shortstack{Number \\of rows}} & \multirow{3}{*}{nz} & \multirow{3}{*}{nnz} & \multicolumn{2}{c|}{CPU time (ms)} & \multicolumn{6}{c|}{Numerical factorization time (ms)}                              \\ \cline{5-12}
                           &                                                &                     &                      & \multirow{2}{*}{GLU3.0}            & \multirow{2}{*}{GLU2.0}  & GLU3.0  & GLU2.0  & \scriptsize{NICSLU}                     & \scriptsize{Speed-up}           & \scriptsize{Speed-up}                 & \scriptsize{Speed-up} \\ 
                           &                                                &                     &                      &                                    &                          & (GPU)   & (GPU)   & \scriptsize{(CPU)~\cite{Chen:TCAD'13}}  & \scriptsize{over GLU2.0}        & \scriptsize{over \cite{Lee:TVLSI'18}} & \scriptsize{over \cite{Chen:TCAD'13}} \\ \hline
rajat12                    & 1879                                           & 12926               & 13948                & 3.999                              & 13.998                   & 2.237   & 2.44883 & 3.99                                    & 1.1                             & 1.0                                   & 1.78 \\ \hline
circuit\_2                 & 4510                                           & 21199               & 32671                & 7.998                              & 59.991                   & 4.144   & 8.36301 & 6.66                                    & 2.0                             & 1.9                                   & 1.61 \\ \hline
memplus                    & 17758                                          & 126150              & 126152               & 15.997                             & 377.943                  & 6.672   & 6.90432 & 26.91                                   & 1.0                             & 0.9                                   & 4.03 \\ \hline
rajat27                    & 20640                                          & 99777               & 143438               & 21.997                             & 404.939                  & 10.539  & 23.8673 & 34.44                                   & 2.3                             & 2.0                                   & 3.27 \\ \hline
onetone2                   & 36057                                          & 227628              & 1306245              & 353.946                            & 36729.4                  & 60.964  & 550.598 & 432.69                                  & 9.0                             & 8.3                                   & 7.10 \\ \hline
rajat15                    & 37261                                          & 443573              & 1697198              & 423.936                            & 18461.2                  & 71.135  & 458.611 & 356.90                                  & 6.4                             & 6.1                                   & 5.02 \\ \hline
rajat26                    & 51032                                          & 249302              & 343497               & 76.988                             & 2011.69                  & 32.366  & 104.12  & 88.77                                   & 3.2                             & 4.2                                   & 2.74 \\ \hline
circuit\_4                 & 80209                                          & 307604              & 438628               & 295.955                            & 4662.29                  & 68.944  & 394.995 & 118.23                                  & 5.7                             & 9.1                                   & 1.71 \\ \hline
rajat20                    & 86916                                          & 605045              & 2204552              & 2190.67                            & 121207                   & 241.822 & 2538.24 & 245.63                                  & 10.5                            & 8.8                                   & 1.02 \\ \hline
ASIC\_100ks                & 99190                                          & 578890              & 3638758              & 2052.69                            & 316998                   & 215.493 & 2652.79 & 357.53                                  & 12.3                            & 14.1                                  & 1.66 \\ \hline
hcircuit                   & 105676                                         & 513072              & 630666               & 67.99                              & 6279.05                  & 46.996  & 243.846 & 221.50                                  & 5.2                             & 9.5                                   & 4.71 \\ \hline
Raj1                       & 263743                                         & 1302464             & 7287722              & 7240.9                             & 140008                   & 845.189 & 7969.05 & 825.38                                  & 9.4                             & 8.7                                   & 0.98 \\ \hline
ASIC\_320ks                & 321671                                         & 1827807             & 4838825              & 2336.64                            & 410679                   & 216.517 & 5632.8  & 765.35                                  & 26.0                            & 21.3                                  & 3.53 \\ \hline
ASIC\_680ks                & 682712                                         & 2329176             & 4957172              & 1747.73                            & 686421                   & 210.697 & 11771.7 & 614.75                                  & 55.9                            & 18.4                                  & 2.92 \\ \hline
G3\_circuit                & 1585478                                        & 4623152             & 36699336             & 9728.52                            & 1764580                  & 878.153 & 38780.9 & 9232.618                                & 44.2                            & 8.2                                   & 10.51 \\ \hline\hline
\multicolumn{9}{|r|}{Arithmetic mean}                            & 13.0        & 7.1 & 3.51 \\ \hline
\multicolumn{9}{|r|}{Geometric mean}                             & 6.7         & 4.8 & 2.81 \\ \hline
\end{tabular}
\end{table*}

\begin{table}[t]
\centering
\caption{Levelization runtimes}
\label{tbl:level}
\begin{tabular}{|l|l|l|l|l|l|} \hline
\multirow{2}{*}{Matrix} & \multicolumn{2}{c|}{Number of levels} & \multicolumn{3}{c|}{Levelization Time (ms)}  \\ \cline{2-6}
                        & GLU2.0 & this work               & GLU2.0  & this work & speed-up              \\ \hline
rajat12                 & 37     & 39                      & 3.048   & 0.035     & 87.1                  \\ \hline
circuit\_2              & 101    & 102                     & 17.187  & 0.074     & 232.3                 \\ \hline
memplus                 & 147    & 147                     & 345.568 & 0.234     & 1476.8                \\ \hline
rajat27                 & 123    & 125                     & 272.216 & 0.32      & 850.7                 \\ \hline
onetone2                & 1213   & 1213                    & 4009.51 & 1.589     & 2523.3                \\ \hline
rajat15                 & 968    & 968                     & 3680.02 & 2.224     & 1654.7                \\ \hline
rajat26                 & 157    & 158                     & 1703.92 & 0.711     & 2396.5                \\ \hline
circuit\_4              & 228    & 229                     & 5053.39 & 0.944     & 5353.2                \\ \hline
rajat20                 & 1216   & 1219                    & 15931.2 & 3.389     & 4700.9                \\ \hline
%\scriptsize{ASIC\_100ks}& 1626   & 1626                    & 36388.8 & 5.301     & 6864.5                \\ \hline
%hcircuit                & 144    & 145                     & 6122.57 & 1.206     & 5076.8                \\ \hline
%Raj1                    & 1594   & 1595                    & 56580.9 & 11.102    & 5096.5                \\ \hline
%\scriptsize{ASIC\_320ks}& 1669   & 1669                    & 168979  & 8.573     & 19710.6               \\ \hline
%\scriptsize{ASIC\_680ks}& 1450   & 1450                    & 530478  & 10.642    & 49847.6               \\ \hline
ASIC\_100ks             & 1626   & 1626                    & 36388.8 & 5.301     & 6864.5                \\ \hline
hcircuit                & 144    & 145                     & 6122.57 & 1.206     & 5076.8                \\ \hline
Raj1                    & 1594   & 1595                    & 56580.9 & 11.102    & 5096.5                \\ \hline
ASIC\_320ks             & 1669   & 1669                    & 168979  & 8.573     & 19710.6               \\ \hline
ASIC\_680ks             & 1450   & 1450                    & 530478  & 10.642    & 49847.6               \\ \hline
G3\_circuit             & 652    & 688                     & 1741860 & 66.508    & 26190.2               \\\hline\hline
\multicolumn{5}{|r|}{Arithmetic mean} & 8804.1 \\ \hline
\multicolumn{5}{|r|}{Geometric mean} & 3145.8 \\ \hline
\end{tabular}
\end{table}

The proposed GLU3.0 is implemented in C++ and CUDA-C, and compiled
with optimization level 3 (-O3). The tests were run on a server
equipped with Intel(R) Xeon(R) CPU E5-2698 v3, 128GB RAM, and NVIDIA
GTX TITAN X (3072 CUDA cores, 12GB GDDR5 memory, Maxwell
architecture). The test matrices come from the University of Florida Sparse
Matrix Collection~\cite{Davis:UFS'11}, and the same ones tested in
\cite{Lee:TVLSI'18} are used for the sake of comparison. Single precision
floating point is used for computation as the Maxwell architecture does not
support atomic operations for double precision. On newer GPU platforms
that allow double atomic operations, the performance of GLU with double
precision is expected on average 30\% slower compared to that of the single
precision version~\cite{Lee:TVLSI'18}.

First we test the new relaxed data dependency detection method proposed in
Section~\ref{sec:dep_alg}. It is applied to the test matrices to perform
levelization. The results of levelization are presented in
Table~\ref{tbl:level}. For the purpose of saving space, more details
of test matrices such as number of non-zeros can be found in
Table~\ref{tbl:kernel}.

From this table, we have two observations. First, the number of
additional levels resulting from the new dependency detection method
are just a few or even zero. As the number of levels is the most
decisive parameter of runtime of the GPU kernel, this means the
proposed new leveling algorithm would have marginal impacts on the
runtime of numerical factorization on GPU. Second, the runtime of
levelization (Algorithm~\ref{alg:dependency}) has improved
dramatically on all test matrices compared to the existing method. The
previous method of levelization used in GLU2.0
(Algorithm~\ref{alg:dependency_2u}) has to explicitly find all
double-U dependency, which has $O(n^3)$ complexity
and thus makes the runtime of preprocessing non-negligible
(compared to the LU factorization time). However, with an average
speed-up ratio of 8804.1 (arithmetic average) or 3145.8 (geometric
mean), the proposed new method is able to reduce the preprocessing
runtime back into the similar time frame of the preprocessing time in
the plain left-looking based method.

Then we test the performance of GPU kernels of GLU3.0 and compare
it with GLU2.0 and NICSLU~\cite{Chen:TCAD'13}, which is a parallel
sparse matrix LU factorization solver based on CPU. 32 threads are
used when testing the performance of NICSLU. The results are
presented in Table~\ref{tbl:kernel}. The speed-up ratio of the
proposed work over \cite{Lee:TVLSI'18} is calculated based on its
reported speed-up ratio against GLU2.0 using the same testing
matrices. The runtime measured includes the time completing memory
copy. CPU time that comprises of preprocessing and symbolic
analysis is compared as well. As can be seen from the table, despite
slightly more levels as reported in Table~\ref{tbl:level}, the
proposed new GPU kernel still demonstrates a steady speedup over the
kernels from GLU2.0 and the improved version from \cite{Lee:TVLSI'18}.
At least $5$x speed-up can be achieved on average. Furthermore, more
significant improvement can be expected when it comes to bigger
matrices, starting from circuit\_4 with a row number of 80209.
The reason is that the computational tasks of small matrices are
so light that the GPU still allows more parallelizable tasks. On the
other hand, when factorizing larger matrices, the limited GPU
computation power will throttle full parallelization in the GLU. In
these cases, the proposed adaptive kernels can utilize the GPU in a
better way so that more parallelism and shorter runtime is achieved.

\begin{table}[t]
\centering
\caption{GPU kernel runtimes without enabling all 3 kernel modes,
    compared to case 1 where small block mode is disabled, and case 2
    where stream mode is disabled.}
\label{tbl:modes}
\begin{tabular}{|l|l|l|l|l|l|l|} \hline
\multirow{2}{*}{Matrix} & \multicolumn{3}{c|}{GPU time (ms)}& \multicolumn{3}{c|}{Level distribution} \\ \cline{2-7}
                        & GLU3.0  & Case 1 & Case 2 & A   & B   & C    \\ \hline
rajat12                 & 2.237   & 2.776  & 2.158  & 2   & 4   & 33   \\ \hline
circuit\_2              & 4.144   & 4.871  & 4.650  & 1   & 10  & 91   \\ \hline
memplus                 & 6.672   & 9.364  & 7.187  & 4   & 3   & 140  \\ \hline
rajat27                 & 10.539  & 13.069 & 10.665 & 6   & 23  & 96   \\ \hline
onetone2                & 60.964  & 66.126 & 173.863& 14  & 33  & 1166 \\ \hline
rajat15                 & 71.135  & 82.677 & 163.947& 11  & 96  & 861  \\ \hline
rajat26                 & 32.366  & 43.697 & 35.330 & 8   & 36  & 114  \\ \hline
circuit\_4              & 68.944  & 170.49 & 103.515& 7   & 9   & 213  \\ \hline
rajat20                 & 241.822 & 571.95 & 1019.12& 11  & 41  & 1167 \\ \hline
ASIC\_100ks             & 215.493 & 246.84 & 1047.78& 13  & 56  & 1557 \\ \hline
hcircuit                & 46.996  & 59.103 & 47.761 & 10  & 14  & 121  \\ \hline
Raj1                    & 845.189 & 2611.12& 2115   & 29  & 223 & 1343 \\ \hline
ASIC\_320ks             & 216.517 & 311.778& 1094.78& 14  & 50  & 1605 \\ \hline
ASIC\_680ks             & 210.697 & 279.784& 721.589& 14  & 55  & 1381 \\ \hline
G3\_circuit             & 878.153 & 783.592& 877.444& 104 & 327 & 257  \\ \hline
\end{tabular}
\end{table}

To further validate the improvement from the proposed three modes of
kernels, another experiment was conducted, where either one of the two
newly proposed modes (small block mode and stream mode) are disabled,
to show the degradation of performance without them. The results are
listed in Table~\ref{tbl:modes}. In case 1, small mode is disabled.
While in case 2, stream mode is disabled. The number of three
different types of levels are also listed. Comparing GLU3.0 with case
1, we can see that small block mode benefits most matrices except
G3\_circuit. Although the number of type A levels is generally small,
small block mode can still lead to decent improvement. The reason of
G3\_circuit being slower without small block mode is probably that
the number of blocks assigned in small block mode is less than optimal
because the limitation of \eqref{eq:memlimit}. In this case, more
warps should be assigned to a block as the total number of blocks is
limited.
Then comparing GLU3.0 with case 2, a more significant improvement can
be seen from stream mode. Furthermore, stream mode tend to benefit all
matrices, as the results of GLU3.0 are either much faster or at worst
equivalent. Especially, the improvement is more significant for large
matrices such as ASIC\_100ks and Raj1.

\begin{figure}[h]
    \centering
    \includegraphics[width=0.9\columnwidth]{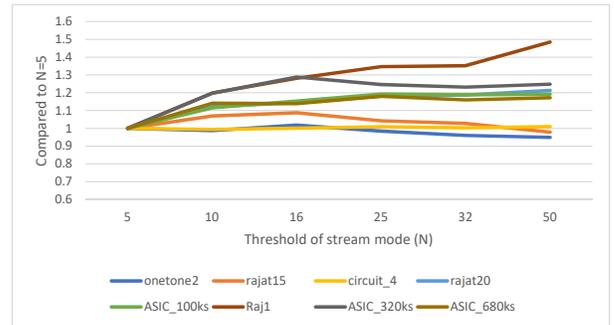}
    \caption{Performance of GPU kernel with different stream mode
    threshold settings}
    \label{fig:threshold}
\end{figure}

In Section~\ref{sec:newkernel}, we mentioned that stream mode starts
when level size decreases to 16. This number is also selected based on
experiment. The results can be found in Figure~\ref{fig:threshold}.
For the purpose of making the figure more clear, instead of using all
matrices used in previous experiments, only the ones that benefit
significantly from stream mode are selected. In the figure, $N$ stands
for the threshold of level size where stream mode begins, and the
values plotted are GPU kernel runtimes with different $N$ compared
with that with $N=5$. It can be seen that the runtime keeps reducing
until $N=16$. Except matrix Raj1, experiments with all other matrices
show slower or equivalent results for larger $N$, which proves that
$N=16$ is a good choice.

According to profiling results, there being unused warps is the
main challenge for this problem. Actually the newly proposed three
modes of kernels have greatly improved utilization of threads in SM,
despite some remaining mismatch due to unpredictable sparsity pattern
of the matrix. This hurts the performance of stream mode most
significantly. As in other modes the warp occupancy is as high as
80\%, while in stream mode the average is 40\%. However, it is also
worth noting that in the ending stage of factorization, as the
submatrix size is decreasing, warp occupancy would drop naturally.

We note that driver overhead is also significant in many of our tests.
Take ASIC\_100ks as an example, the first cuda function call
(including invisible set-up works) takes 40\% of all GPU time (215ms).
For larger matrices, this problem should be less severe for larger
matrices or in real simulation scenarios where the factorization
kernel is called repeatedly.

\section{Conclusion}
\label{sec:conclusion}

We have proposed a new sparse LU solver on GPU for circuit simulation
and more general scientific computing. The new sparse LU solver,
called {\it GLU3.0} has two main improvements: First 
a more efficient data dependency detection algorithm is introduced.
Second, three different kernel operation modes are developed based on
different number of columns in a level, as the LU factorization
progresses. They enable dynamic allocation of GPU blocks
to better balance the computing demands and resources during the LU
factorization process. Numerical results on the set of typical
circuit matrices from University of Florida Sparse Matrix Collection
(UFL) have shown that GLU3.0 can deliver 2-3 orders of magnitude
speedup over GLU2.0 for the data dependency detection. Furthermore,
GLU3.0 achieves 13.0 $\times$ (arithmetic mean) or 6.7 $\times$
(geometric mean) speedup over GLU2.0 and 7.1 $\times$ (arithmetic
mean) or 4.8 $\times$ (geometric mean) over recent proposed enhanced
GLU2.0 sparse LU solver on the same set of circuit matrices.

% references section
%\bibliographystyle{ACM-Reference-Format}
%\begin{thebibliography}{9}
\footnotesize
\bibliographystyle{unsrt}
\bibliography{../../bib/softerror,../../bib/stochastic,../../bib/simulation,../../bib/modeling,../../bib/reduction,../../bib/misc,../../bib/architecture,../../bib/mscad_pub,../../bib/thermal_power}
%\end{thebibliography}

%\vspace{-0.7in}
\begin{IEEEbiography}
  [{\includegraphics[width=1in,height=1.25in,clip,keepaspectratio]{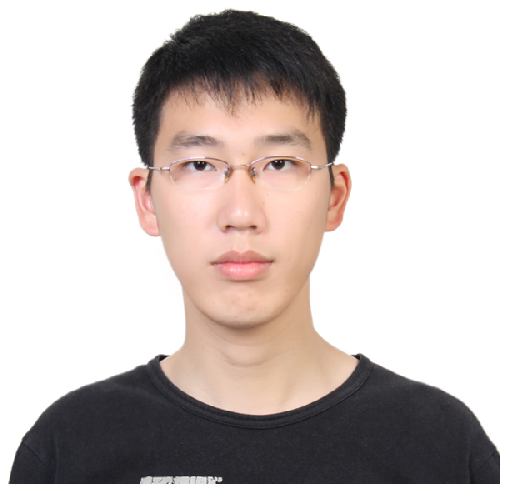}}]
  {Shaoyi Peng} (S'17) received his B.S degree in Microelectronics
  from Fudan University, Shanghai, China in 2016. He is currently a
  Ph.D. candidate in the Department of Electrical and Computer
  Engineering at the University of California, Riverside. His research
  interest includes VLSI reliability effect modeling and simulation,
  finite element method analysis and numerical methods.
\end{IEEEbiography}
\vspace{-0.6in}

\begin{IEEEbiography}
   [{\includegraphics[width=1in,height=1.25in,clip,keepaspectratio]{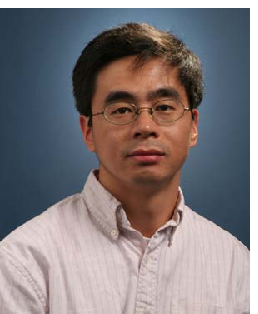}}]
{Sheldon X.-D. Tan} (S'96-M'99-SM'06) received his B.S. and M.S.
  degrees in electrical engineering from Fudan University, Shanghai,
  China in 1992 and 1995, respectively and the Ph.D. degree in
  electrical and computer engineering from the University of Iowa,
  Iowa City, in 1999. He is a Professor in the Department of
  Electrical Engineering, University of California, Riverside, CA. He
  also is a cooperative faculty member in the Department of Computer
  Science and Engineering at UCR. His research interests include VLSI
  reliability modeling, optimization and management at circuit and
  system levels, hardware security, thermal modeling, optimization and
  dynamic thermal management for many-core processors, parallel
  computing and adiabatic and Ising computing based on GPU and
  multicore systems. He has published more than 300 technical papers
  and has co-authored 6 books on those areas.

  Dr. Tan received NSF CAREER Award in 2004. He also received three
  Best Paper Awards from ICSICT'18, ASICON’17, ICCD'07, DAC’09. He
  also received the Honorable Mention Best Paper Award from
  SMACD’18. He was a Visiting Professor of Kyoto University as a JSPS
  Fellow from Dec.  2017 to Jan. 2018. He is serving as the TPC Chair
  of ASPDAC 2021, amd TPC Vice Chair of ASPDAC 2020.  He is serving or
  served as Editor in Chief for Elsevier’s Integration, the VLSI
  Journal, the Associate Editor for three journals: IEEE Transaction
  on VLSI Systems (TVLSI), ACM Transaction on Design Automation of
  Electronic Systems (TODAES) and Elsevier’s Microelectronics
  Reliability.
\end{IEEEbiography}

\end{document}